# Emergence of half-semiconductor behavior and tunable magnetism via carrier doping in Janus VXSe (X=Cl, Br, I) monolayers

Zhixiang Wang,[1,2#] Aining Wang,[1,#] Zikui Ye,[3] Yuqiao Qu,[4] Qian Zheng,[1] Yiding Liu,[5] Qiang Fan,[6] and Gang Yao[1,7,*]

[1]School of Physical Science and Technology, Southwest University, Chongqing 400715, China

[2]Hanhong College, Southwest University, Chongqing 400715, China

[3]AECC Hunan Aviation Powerplant Research Institute, Zhuzhou 412002, Hunan, China

[4]Department of Physics, Chengdu University of Technology, Chengdu 610059, China

[5]College of Physics and Optoelectronic Engineering, Leshan Normal University, Leshan 614004, China

[6]School of New Energy Materials and Chemistry, Leshan Normal University, Leshan 614004, China

[7]Tsung-Dao Lee Institute, Shanghai Jiao Tong University, Shanghai 200240, China

[#]These authors contributed equally to this work.

*To whom correspondence should be addressed. Email: yaogang@swu.edu.cn

**Two-dimensional ferromagnetic semiconductors with high Curie temperature, large magnetic anisotropy, and electrically tunable properties are highly sought for nanoscale spintronics. Here, using first-principles calculations, we predict a new class of Janus VXSe (X = Cl, Br, I) monolayers. All three compounds are intrinsic ferromagnetic semiconductors with indirect band gaps of 1.66-2.33 eV, Curie temperatures up to 128 K, and magnetic anisotropy energies up to 910 μeV per V, leading to easy-plane magnetization for VClSe and VBrSe and an out-of-plane easy axis for VISe. The robust ferromagnetic order originates from the competition between superexchange and itinerant magnetism. Remarkably, VISe is a pristine half-semiconductor, whereas carrier doping unlocks fully spin-polarized states in the initially non-ideal VClSe and VBrSe. We also find that hole doping can switch the magnetic easy axis of VBrSe from in-plane to out-of-plane, enabling a spin-field-effect transistor with giant magnetoresistance. Our findings highlight carrier doping as a key to unlock hidden half-semiconducting behavior and establish Janus VXSe as a promising platform for 2D spintronics and magnetoelectric devices.**

Spintronics exploits the electron spin for information storage, processing, and transport, attracting intensive interest from science and industry.[1-3] Ferromagnetic (FM) crystals with large band gap, high Curie temperature ($T_c$), and large magnetic anisotropy are fundamental for spintronic devices.[4-6] Realizing this at the nanoscale requires low-dimensional ferromagnetic semiconductors (FMSs) with these qualities. Two-dimensional (2D) materials have emerged as a fertile ground. Graphene offers long spin diffusion but lacks intrinsic magnetism;[7] transition metal dichalcogenides like $MoS_2$ and Janus MoSSe show strong spin-orbit coupling (SOC) yet remain nonmagnetic in their pristine form.[8, 9] Over the years, for example, in monolayer 2D $MoS_2$, a variety of extrinsic approaches have been tried, including magnetic doping (Fe-doping induced room-temperature ferromagnetism)[10], strain engineering,[11, 12] and magnetic proximity effects[13]). Ferromagnetism can also be induced in graphene nanoribbons by electric fields or magnetic defects, and under carrier doping, nonmagnetic 2D semiconductors such as GaSe[14] and $\alpha$-XO (X=Sn, Pb)[15, 16] become FM. Despite these efforts, achieving stable, uniform magnetic order in 2D remains challenging. Recently, exfoliated van der Waals crystals offered an intrinsic alternative: semiconducting $CrI_3$ ($T_c$=45 K)[17] and $Cr_2Ge_2Te_6$ ($T_c$=20 K)[18] prove that FM order survives to atomic thin films, but their $T_c$ lies far below 77 K, limiting practical applications.

On the other hand, two additional functionalities are highly desired for next-generation spintronics: 100% spin-polarized carriers and electrically switchable magnetic easy axis. First, half-semiconductors (HSCs), where valence and conduction band edges reside in the same spin channel, provide fully spin-polarized carriers upon injection (*e.g.*, by electrical gating),[19] a property demonstrated in 3D bulk compounds like $La_2NiMnO_4$[20] and predicted in monolayers such as VOF,[21] strained-$ZrCl_3$,[22] and doped $SnSe_2$.[23] Various approaches (strain, doping) have been explored to achieve 100% spin-polarization, including bipolar magnetic semiconductors (BMSs).[24-27] HSCs instead offer a robust, concentration-independent source of spin-polarized carriers. Yet a key question remains: can a material without an ideal HSC ground state still exhibit the hallmark of an HSC? Second, electrical control of magnetic anisotropy, essential for low-power non-volatile memory, has been realized in only a few 2D magnets (*e.g.*, by electrostatic gating of $Cr_2Ge_2Te_6$[28] and by ionic-liquid gating of $Fe_5GeTe_2$,[29] or proposed in some monolayers.[30,31] Until now, such switchable anisotropy remains extremely rare, and the effect of carrier doping on the ground state has received little theoretical attention.

Herein, by means of first-principles calculations, ab initio molecular dynamics, and Monte Carlo simulations (see Supplemental Material for computational details), we investigate the electronic and magnetic properties of Janus VXSe (X = Cl, Br, I) monolayers. We find that VISe is a pristine FM HSC with both band edges residing in the same spin channel. VClSe and VBrSe , in contrast, are initially non-ideal (only the CBM spin-polarized) but become fully spin-polarized upon carrier doping, revealing a doping-induced emergence of half-semiconductivity. Their FM order arises from a competition between superexchange and itinerant magnetism. Moreover, hole doping switches the magnetic easy axis of VBrSe from in-plane to out-of-plane, driven by a sign reversal of the Se orbital contribution to the MAE. The microscopic mechanism for the origin of electrically magnetism is also discussed. These results extend the HSC paradigm beyond pristine systems and suggest that 2D VXSe are promising materials for 2D spintronics, such as spin valves and spin-field-effect transistor (spin-FETs).

Figures 1(a) and 1(b) display the structures of the 2D VXSe (X = Cl, Br, I) monolayers. They form a hexagonal lattice with space group P3-M1 (No. 156). Each V ion sits in a distorted octahedron coordinated by three halogen and three Se atoms, analogous to the 1T phase of transition metal dichalcogenides. After full relaxation (Table S1), the lattice constants follow the ionic size trend: $r_{Cl} < r_{Br} < r_{I}$. The V-X bond lengths increase from Cl to I, all longer than V-Se. The bond angles around V (X-V-X, X-V-Se, and Se-V-Se) are all close to 90°, with slight variations depending on the halogen. This deviation from perfect octahedral symmetry, combined with the Janus-induced vertical asymmetry, creates a distorted ligand field affecting the magnetic anisotropy. Phonon dispersion and AIMD simulations confirm dynamical and thermal stability at 300 K (Fig. S1).

Magnetic calculations identify the FM ground state for all three crastals (Fig. S2). The net moment (~2.2 $\mu_B$ per V) originates mainly from the V atoms; the halogen and Se atoms carry small induced moments (~-0.2 $\mu_B$) antiparallel to V. This is also confirmed by the spin-polarized charge density in Figs. 1(e) and 1(f), indicating substantial spin polarization. Because GGA+U severely underestimates band gaps and can even misrepresent the spin character of band edges in correlated systems,[32,33] we use the HSE06 hybrid functional for all quantitative conclusions. The HSE06 results show that all three are FMSs with indirect band gaps of 2.15 eV (VClSe), 2.33 eV (VBrSe), and 1.66 eV (VISe) [Figs. 1(c),1(d), and S4]. Notably, VISe is a pristine HSC (both band edges in spin-up channel), whereas VClSe and VBrSe are not, only their conduction band minimum is spin-polarized. This is in contrast to the GGA+$U$ results (Fig. S3), which incorrectly predict all three to be HSCs due to the well-known over-delocalization error of (semi)local functionals. Therefore, the reliable conclusion from the more accurate HSE06 functional is that carrier doping is essential to unlock half-semiconducting behavior in VClSe and VBrSe, a point we return to later.

The stability of the FM ground state at finite temperatures is governed by the magnetic exchange interactions. We extract the magnetic exchange couplings up to the third nearest neighbor using a classical Heisenberg model. The spin Hamiltonian is

$$H = -J_1 \sum_{i,j} S_i S_j - J_2 \sum_{i,j} S_i S_j - J_{i,j} \sum_3 S_i S_j, \quad (1)$$

with $|S_i|=1$. By mapping DFT total energies of four magnetic configurations to this Hamiltonian,we obtained the exchange parameters (Table S2). The nearest-neighbor coupling $J_1$ is found to be positive and dominant, establishing the FM ground state. $J_2$ is also positive but an order of magnitude weaker, whereas $J_3$ is negative (AFM). Among the three materials, VISe has the largest $J_1$ and the smallest $|J_3|$.

The origin of the FM coupling follows from the electronic configuration and geometry. Two exchange mechanisms compete. The dominate FM $J_1$ originates from a strong dual-path superexchange ($J_1^{\mathrm{FM}}$) and a weak V-V FM direct exchange ($J_1^{\mathrm{AFM}}$) mechanism. In an ideal octahedral field, the V $3d$ orbitals split into a lower-lying half-occupied triplet $t_{2g}$ and a higher-lying empty $e_g$ states. The Janus structure, with distinct X and Se ligands, breaks inversion symmetry and further lifts orbital degeneracy. As illustrated in Figs. 2(a) and 2(b), with an $d^2$, specifically the $t_{2g}^2 e_g^0$, electronic configuration for the $V^{2+}$ ion, the near-orthogonal $t_{2g}^2 - p - e_g^0$ pathways support strong FM superexchange interaction through both anion layers, in line with the Goodenough-Kanamori-Anderson (GKA) rules for nearly 90° bond angles.[34-36] Based on a general band coupling model,[37-39] level repulsion between partially occupied $t_{2g}$ states also favors FM direct exchange between nearest $V^{2+}$ ions, but this contribution is weak and negligible because of the large V-V distance (~3.6 Å). As for $J_2$, which involves V-X-Se-V paths; their geometry allows multiple superexchange channels whose net sign depends on bond angles and orbital overlaps. $J_3$ arises from longer paths (*e.g.*, V-Br-Se-Br-V). The ratio $|J_3|/|J_2|$ shows strong halogen dependence: 3.0 for Cl, 1.0 for Br, and 0.16 for I, reflecting sensitive competition between FM and AFM contributions. The small magnitude of $J_3$ in all cases is consistent with its long-range origin.

Magnetic anisotropy energy (MAE) , the energy difference between in-plane and out-of-plane spin configurations, determines both the preferred magnetization orientation and the robustness against thermal fluctuations.[40,41] We obtain MAE values of -363 μeV (VClSe), -243 μeV (VBrSe), and 910 μeV (VISe) per V atom, where positive (negative) denotes an out-of-plane (in-plane) magnetization. The 3D MAE maps in Figs. 2(c) and 2(d) confirm that VClSe and VBrSe are easy-plane magnets with isotropic magnetic polarization, placing them in the class of 2D XY magnets, similar to monolayer CrCl.[42] In contrast, VISe exhibits perpendicular magnetic anisotropy, consistent with the Ising-type ferromagnetism observed in several experimental 2D magnets.[17,18,43,44] The MAE magnitudes in VXSe are substantially larger than those of cubic metals (below 70 μeV/atom) and comparable to those of Fe and Co monolayers on Rh(111) and Pt(111) substrates (0.08-0.37 meV/atom).[45] These large anisotropy energies make VXSe promising for magnetoelectronic applications. Using Wolff Monte Carlo simulations based on Eq. (1) with the magnetic anisotropy term included, we predict $T_c$ of 93 K for VClSe, 100 K for VBrSe, and 128 K for VISe, respectively [Fig. 2(e)]. The highest value for VISe reflects enhanced *d-p* hybridization as the halogen electronegativity decreases. We validate our protocol on monolayer $CrI_3$, obtaining $T_c$=50K,[46] consistent with the experimental value of 45 K.[17]

Although they are lower than those of high-performance Janus magnets such as CrXY (X=S, Se, Te, Y=F, Cl, Br, I; $E_g$~2 eV, $T_c$ ~300 K)[47] and $V_2X_3I_3$ (X=Cl, Br, I; $E_g$=2.3 eV and $T_c$~240 K),[48] VXSe maintain considerably wider band gaps (up to 2.33 eV). Within intrinsic FMs, their $T_c$ compare favorably with many established systems (typically 12-82 K).[24,25,49-52] For instance, most reported FMSs with higher $T_c$ suffer from much narrower gaps, for example, $LiFe_2X_2$ (X=S, Se, Te; 35 meV and 1500 K),[53] TbXH (X=Br, Cl, I; 0.5 eV, 113 K),[54] and VSeX (X=S, Te; 0.04-0.13 eV and 82-106 K),[55] etc.,[56,57] whereas wide-gap systems like $CoBr_2$ (2.36 eV, 1.7 K) exhibit the opposite trade-off. This inverse relationship between exchange strength and band gap reflects a general trend in correlated electron systems: enhancing ferromagnetic exchange via increased *d/f-p* hybridization often reduces the charge transfer energy, which tends to narrow the electronic gap.[55,58]

The reduced magnetic moment in VXSe suggests partial 3*d* electrons delocalization, pointing to itinerant character. Combined with the half-semiconducting band structures, this electronic configuration raises the possibility that carrier injection could enable not only fully spin-polarized currents but also electrically tunable magnetism, a prospect we now test via controlled doping. The maximum charge doping considered here is about $2.5\times10^{14}/cm^2$ (~0.1*e* per atom), experimentally accessible. For instance, carrier densities up to $10^{15}/cm^2$ have been achieved in 2D systems using ionic liquid gating, as demonstrated in gated ZnO and $La_{0.8}Ca_{0.2}MnO_3$ transistors.[59,60]

Let us first examine the band structure under carrier doping, taking VBrSe as an example. As displayed in Fig 3(a), unexpectedly, we found that in both the hole and electron doping cases, the system can become half-metallic, with the $E_F$ crossing a fully spin-polarized band. The conducting spin channel remains the same regardless of doping type, confirming spin polarization locked to band edges. Below a threshold concentration ($\pm1\times10^{14}/cm^2$), the material retains semiconducting. This distinguishes HSCs from conventional half-metals and BMSs, where 100% spin polarization requires precise $E_F$ tuning and the polarization direction may invert with doping type.[24-26,30] Thus, despite its non-ideal pristine state, VBrSe becomes a HSC under carrier doping, extending the HSC paradigm to materials that are half-semiconducting only under carrier doping.

The magnetic ground state is sensitive to carrier concentration, which in turn modifies the band structure. To explore this effect, we computed the total energies of different magnetic configurations as a function of carrier concentration [Fig. 3(b)] and found that hole doping systematically reduces the energy difference. At the highest hole concentration, the $AFM_1$ state becomes lower in energy at 0 K. However, static energy comparisons miss finite-temperature effects.[61,62] Monte Carlo simulations confirm that the FM phase remains robust, with a nonzero $T_c$ even at this doping level [Fig. 3(c)]. The doping dependence of $T_c$ and the local moment $M_V$ reveals coupling between magnetic moments and itinerant carriers. Hole doping reduces both quantities, while electron doping enhances them, an asymmetry consistent with strengthened exchange interactions from injected electrons. This behavior aligns with the trend of enhanced *d-p* hybridization and charge transfer across the halogen series, where heavier halogens promote itinerant exchange.

Electrical control of magnetic anisotropy is highly desired for magnetoelectric applications.[28,29,31] Figure 4(a) shows the energy difference between in-plane and out-of-plane magnetization as a function of carrier concentration in VBrSe. At low hole doping, the material favors in-plane magnetization. As hole density increases, in-plane anisotropy weakens, and a spin reorientation transition occurs at a critical concentration of $-1.5\times10^{14}/cm^2$. Beyond this threshold, out-of-plane magnetization becomes energetically favored, with the anisotropy energy reaching ~110 μeV at the highest hole concentration. In contrast, electron doping only modestly enhances the in-plane anisotropy without inducing a transition. The underlying mechanism is discussed in the supplemental Material.

This electrically controlled easy axis switching enables a spin-FET.[63] As illustrated in Fig. 4(b), a double-gated structure with dielectric layers (*e.g.*, *h*-BN and $MoS_2$) allows independent carrier density control in adjacent regions of the VBrSe channel. When a gate voltage induces a hole density exceeding $n_c$ in a local region, the easy axis switches from in-plane to out-of-plane, while the undoped region retains in-plane magnetization, creating a lateral heteromagnetic interface with orthogonal alignment. Spin-dependent scattering at this interface produces a high-resistance state, whereas a uniformly magnetized channel yields a low-resistance state. Thus, a gate-tunable giant magnetoresistance effect is realized through electrostatic doping. Notably, the hole concentration required to achieve half-metallicity is lower than that for MAE switching, so the half-metallic state already exists when the easy axis flips, which is expected to significantly enhance the magnetoresistance ratio relative to conventional ferromagnetic semiconductor channels.

In addition to VBrSe, we also investigated VClSe and VISe under carrier doping [see Figs. S5, S6, and S8). Under doping, VClSe behaves similarly to VBrSe, becoming half-semiconducting, while VISe does so intrinsically. Both retain their FM ground state and show no qualitative change in MAE under doping. The $T_c$, however, responds to doping in opposite ways: in VClSe, both electrons and holes lower $T_c$; in VISe, holes raise $T_c$ while electrons lower it. Adding carriers always increases the local moment, a trend observed in all three compounds due to Coulomb screening and band filling, signaling itinerant electron involvement. The net change in $T_c$ depends on which of the two exchange contributions prevails: superexchange weakens with mobile carriers, whereas itinerant exchange strengthens. Moving from Cl to I, the orbital overlap between V and the anion increases and the electronegativity difference decreases, both of which enhance itinerant exchange. Thus, in VClSe superexchange remains dominant, so any added carrier lowers $T_c$; in VISe the heavier iodine boosts itinerant exchange enough that hole doping overcomes the superexchange loss and raises $T_c$. VBrSe sits in between: both doping types suppress $T_c$, but holes also flip the easy axis. These results provide a coherent picture of how the halogen controls the magnetic response across the VXSe family. They also show that 2D VXSe are a rare system where one can tune the competition between localized superexchange and itinerant exchange by halogen substitution or by applying a gate voltage.

To summarize, we presented a new family of 2D Janus VXSe (X = Cl, Br, I) monolayers. Spin-polarized calculations with the hybrid HSE06 functional reveal that all three are FMSs with wide band gaps (up to 2.33 eV) and sizable MAE, with the latter being one to two orders larger than those of Co, Fe, and Ni bulk. Using the Heisenberg model, we predict $T_c$'s of 93 K (VClSe), 100 K (VBrSe), and 128 K (VISe). Significantly, VISe is suggested to be a pristine HSC, whereas VClSe and VBrSe become half-semiconducting only under carrier doping, thereby extending the HSC paradigm beyond pristine systems. Electrostatic doping further switches the easy axis of VBrSe from in-plane to out-of-plane, enabling a spin-FET. Although the magnetic moments on V atoms couple ferromagnetically via superexchange, the distinct evolution of $T_c$ under doping (suppressed in VClSe by both carriers but enhanced by holes in VISe) reveals a competition between superexchange and itinerant magnetism. These results establish Janus VXSe as a versatile platform where intrinsic ferromagnetism, wide band gaps, and electrically tunable magnetism coexist, revealing a competition between superexchange and itinerant exchange that can be controlled by halogen substitution and carrier doping. This mechanistic understanding, together with the electrically switchable half-semiconducting behavior and magnetic easy axis, positions VXSe monolayers as a compelling system for exploring doping-controlled magnetism and for potential applications in 2D spintronic devices.

**Figures and figure captions:**

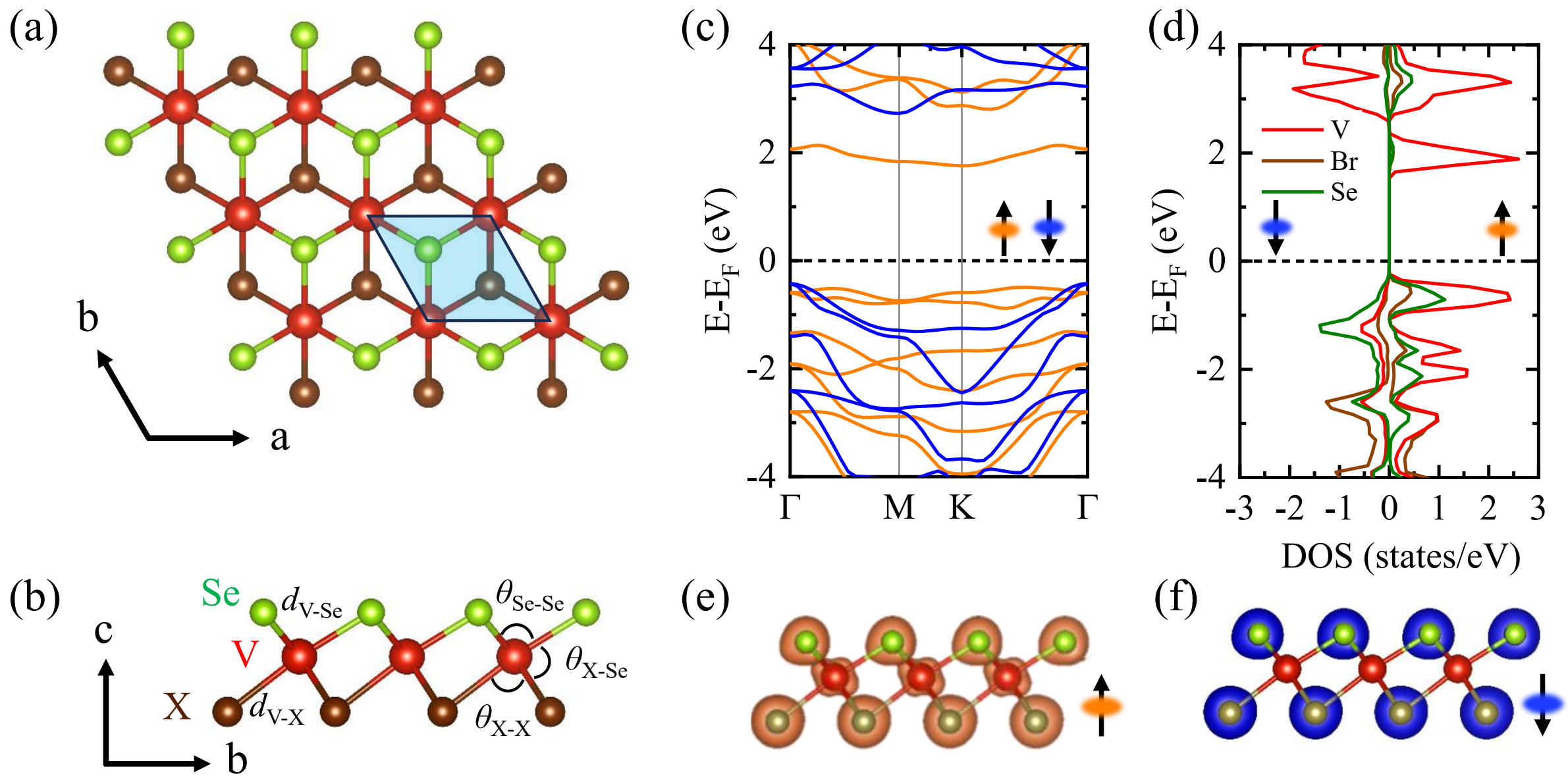

**FIG. 1.** Monolayer VXSe and their electronic properties. (a) Top and (b) side views of the VXSe monolayers. In panel (a), the blue shaded area indicates the primitive cell. In panel (b), the X-V-Se, X-V-X, and Se-V-Se bond angles are indicated as $\theta_{X\text{-}Se}$, $\theta_{X\text{-}X}$ and $\theta_{Se\text{-}Se}$, respectively. The distances between V and X (Se) atoms are indicated by $d_{V\text{-}X}$ ($d_{V\text{-}Se}$). (c) Spin-polarized band structures and (d) projected density of states of the VBrSe monolayer. (e) and (f) Spin-dependent charge densities for VBrSe.

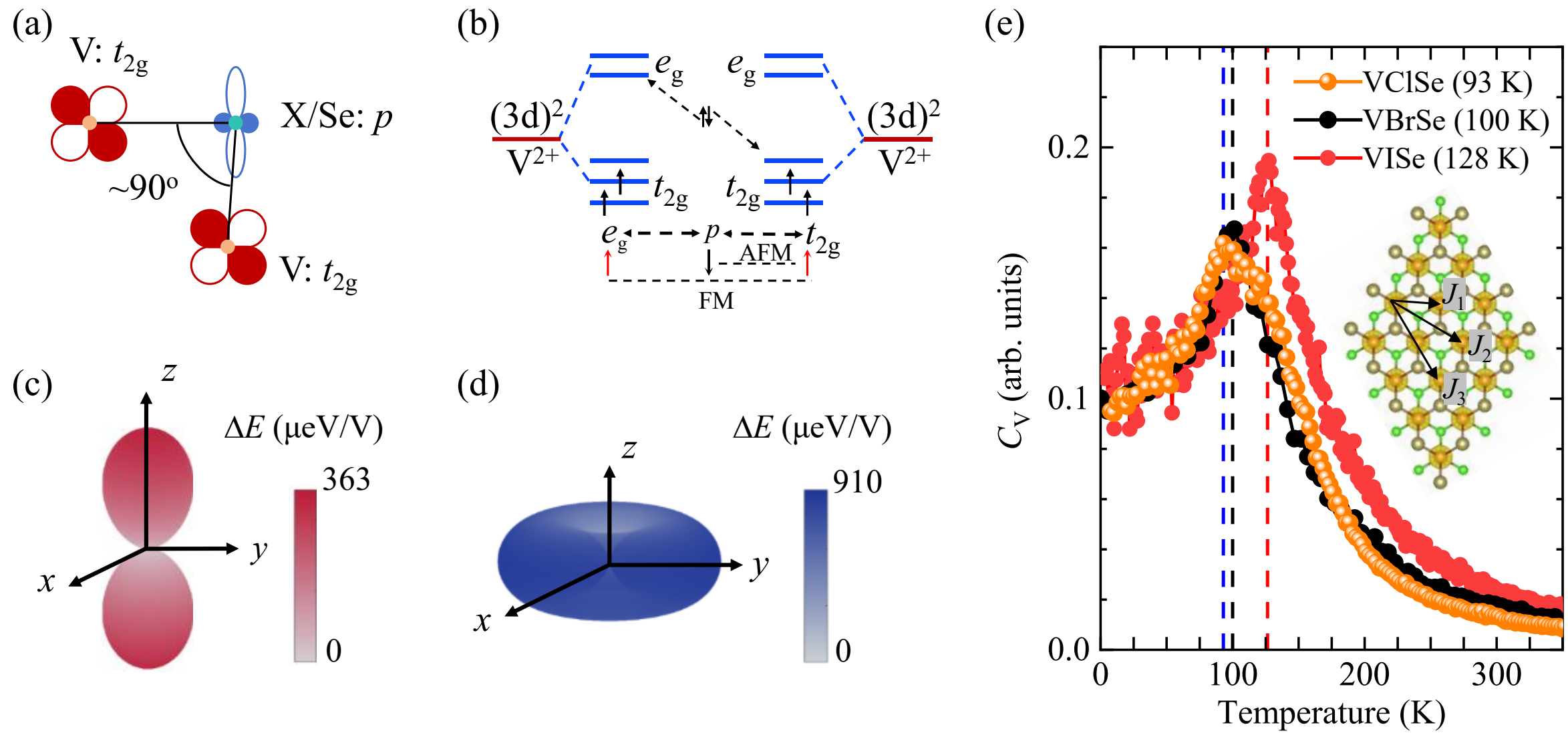

**FIG. 2.** Magnetic properties. (a) and (b) Schematic illustrations of the FM superexchange interaction in VXSe monolayers via the $t_{2g}$-$p$-$e_g$ orbital pathways. 3D MAE maps for (c) VBrSe and (d) VISe. The in-plane FM state is set as the energy reference (zero) for VBrSe (and similarly for VClSe, not shown), while the out-of-plane FM state is used as the reference for VISe. The radial distance and color saturation of each point represent the relative energy increase required to rotate the magnetic moment to that orientation in 3D space. (e) Specific heat $C_V$ as a function of temperature. The inset shows the spin-polarization distribution and the exchange paths for nearest ($J_1$), next-nearest ($J_2$), and third-nearest ($J_3$) neighbor exchange interactions.

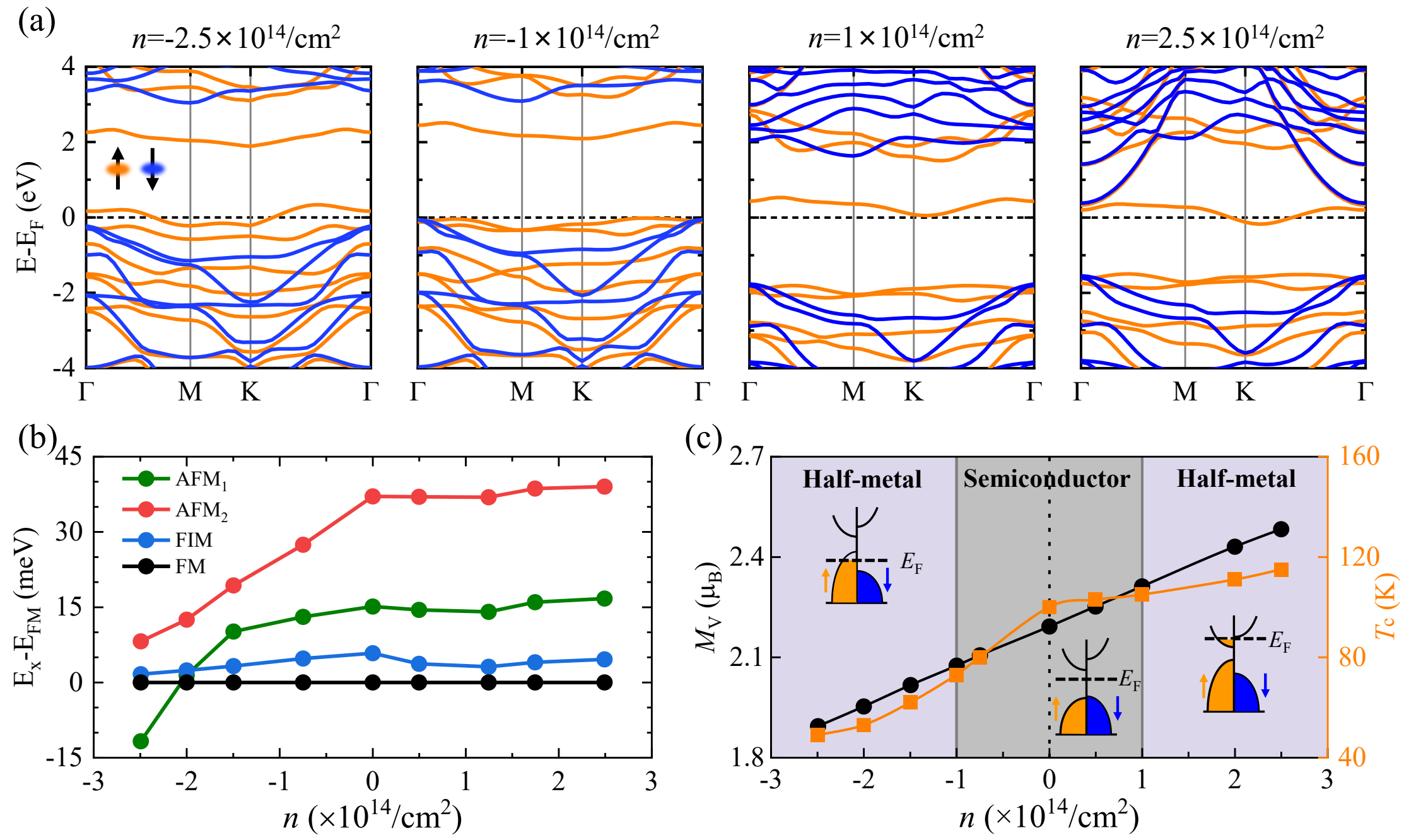

**FIG. 3.** Carrier doping-induced half-metallicity feature in VBrSe. (a) Band structures for VBrSe at carrier concentrations (from left to right) of $n$=-2.5×10$^{14}$, -1×10$^{14}$, 1×10$^{14}$, and 2.5×10$^{14}$/cm$^2$; negative and positive values correspond to hole and electron carrier doping, respectively. (b) Total energies of various magnetic configuration for monolayer VBrSe as a function of carrier density. (c) Phase diagram under carrier doping. The data points indicate the local magnetic moment on V site and $T_c$. Shaded regions distinguish semiconductor and half-metal (HM) regimes. Insets show the corresponding schematic plots of band structures.

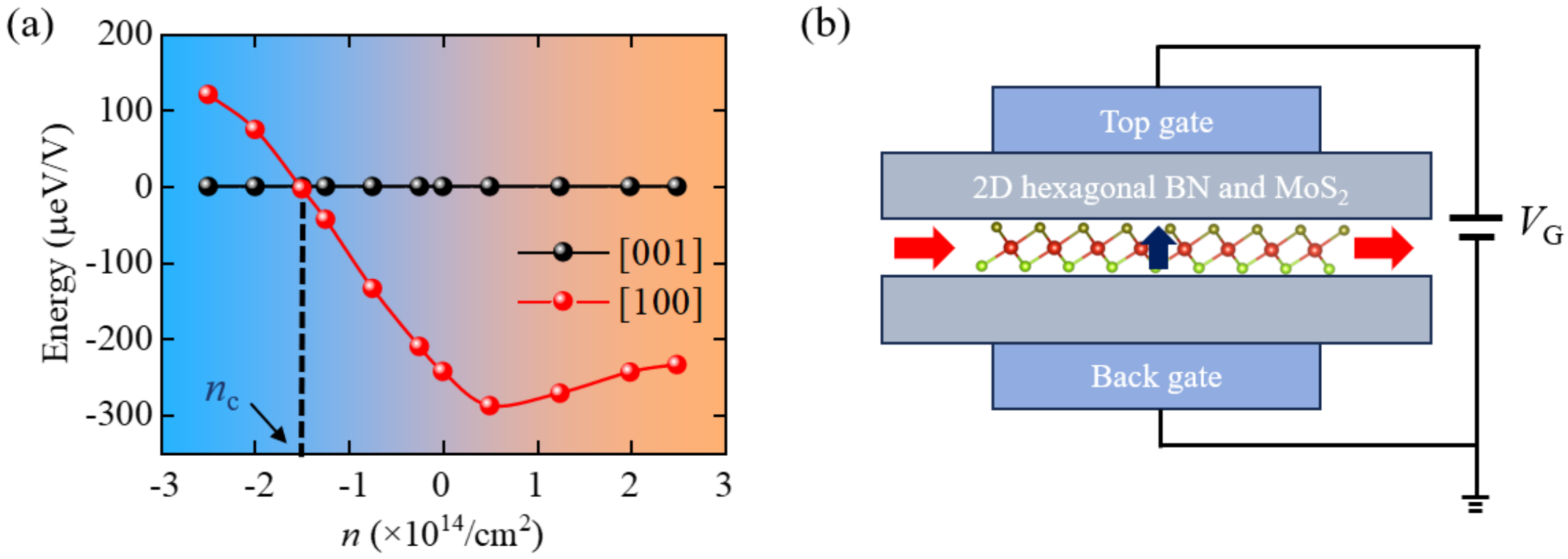

**FIG. 4.** Switchable easy axis in VBSe. (a) Energy of the FM state for monolayer VBrSe with different magnetization directions as a function of carrier concentration. (b) Schematic of a double-gated 2D spin-FET. The gate voltage controls the carrier concentration, concurrently switching the magnetic easy axis and driving a semiconductor-to-half-metal transition, which enables a gate-tunable magnetoresistance effect.